\let\dl=\bf

\catcode`\@=11

\newif\if@fewtab\@fewtabtrue
{\count255=\time\divide\count255 by 60
\xdef\hourmin{\number\count255}
\multiply\count255 by-60\advance\count255 by\time
\xdef\hourmin{\hourmin:\ifnum\count255<10 0\fi\the\count255}}
\def\ps@draft{\let\@mkboth\@gobbletwo
    \def\@oddhead{}
    \def\@oddfoot{\hbox to 7 cm{\tiny \versionno
       \hfil}\hskip -7cm\hfil\rm\thepage \hfil}
    \def\@evenhead{}\let\@evenfoot\@oddfoot}

\def\draftcite#1{\ifnum\draftcontrol=1#1\else{}\fi}
\def\@lbibitem[#1]#2{\item{}\hskip -3cm \hbox to 2cm
{\hfil$\scriptstyle\draftcite{#2}$}\hskip
1cm[\@biblabel{#1}]\if@filesw
     {\def\protect##1{\string ##1\space}\immediate
      \write\@auxout{\string\bibcite{#2}{#1}}}\fi\ignorespaces}

\def\@bibitem#1{\item\hskip -3cm \hbox to 2cm
{\hfil {\footnotesize\draftcite{#1}}}\hskip 1cm
\if@filesw \immediate\write\@auxout
       {\string\bibcite{#1}{\the\value{\@listctr}}}\fi\ignorespaces}

\def\citen#1{\if@filesw \immediate\write \@auxout {\string\citation{#1}}\fi%
\@tempcntb\m@ne \let\@h@ld\relax \def\@citea{}%
\@for \@citeb:=#1\do {\@ifundefined {b@\@citeb}%
    {\@h@ld\@citea\@tempcntb\m@ne{\bf ?}%
    \@warning {Citation `\@citeb ' on page \thepage \space undefined}}%
    {\@tempcnta\@tempcntb \advance\@tempcnta\@ne
    \setbox\z@\hbox\bgroup\ifcat0\csname b@\@citeb \endcsname \relax
    \egroup \@tempcntb\number\csname b@\@citeb \endcsname \relax
    \else \egroup \@tempcntb\m@ne \fi \ifnum\@tempcnta=\@tempcntb
    \ifx\@h@ld\relax \edef \@h@ld{\@citea\csname b@\@citeb\endcsname}%
    \else \edef\@h@ld{\hbox{--}\penalty\@highpenalty
    \csname b@\@citeb\endcsname}\fi
    \else \@h@ld\@citea\csname b@\@citeb \endcsname \let\@h@ld\relax \fi}%
\def\@citea{,\penalty\@highpenalty\hskip.13em plus.13em minus.13em}}\@h@ld}
\def\@citex[#1]#2{\@cite{\citen{#2}}{#1}}%
\def\@cite#1#2{\leavevmode\unskip\ifnum\lastpenalty=\z@\penalty\@highpenalty\fi%
  \ [{\multiply\@highpenalty 3 #1%
  \if@tempswa,\penalty\@highpenalty\ #2\fi}]}   %
\makeatother % end of CITE.STY

\def\aff           {{\cal A}}
\def\affO          {\brev{\cal A}}

\def\AO            {\brev{A}}

\def\aw            {a_{w_0}}

\def\barW          {\bar W}
\def\bearl         {\begin{array}{l}}
\def\bearll        {\begin{array}{ll}}
\def\bearlll       {\begin{array}{lll}}
\def\bearllll      {\begin{array}{llll}}

\def\be            {\begin{equation}}

\def\bfe           {{\bf1}}
\def\bfem          {{\bf1}}
\def\block         {{\cal B}^{\bfe,k}_G}

\def\brev          {\breve}
\def\caij          {\brev{a}_{ij}}

\def\calf          {{\cal F}}

\def\calH          {{\cal H}}
\def\calm          {{\cal M}}
\def\calmg         {{\cal M}_{\Gm}}
\def\calmomega     {{\cal M}_{\Gm}^\omega}

\def\cft           {conformal field theory}
\def\cfts          {conformal field theories}

\def\Chi           {\chi}

\def\Chil          {\Chi_\Lambda^{[\omega]}}

\def\ChiO          {\brev\Chi}

\def\cI{\breve{I}}

\def\complex       {{\dl C}}

\def\CS            {{\cal S}}

\def\csa           {Cartan subalgebra}
\def\CST           {Chern\hy Simons theory}
\def\CSTs          {Chern\hy Simons theories}

\def\dyd           {Dynkin diagram}
\def\ee            {\end{equation}}
\def\eE            {{\rm e}}

\def\eear          {\end{array}}

\newcommand\erf[1]{(\ref{#1})}

\newcommand\fcft[3]{{{#1}^{\mskip-#3 mu\raise #2 pt\hbox{${\scriptstyle\circ}$}}}}
\def\findim        {finite-dimensional}
\newcommand{\fline}[1]{\vfill\noindent ------------------\\[1 mm]}

\def\futnot#1      {\ifnum\draftcontrol=1
                   \footnote{~{\sc internal footnote:} #1}\ \fi}
\def\futnote#1     {\footnote{~#1}\ }
\def\Gg             {\Gu}
\def\g             {{\sf g}}
\def\gbar          {\bar {\sf g}}
\def\G             {\mbox{$G$}}
\def\Gu            {\mbox{$\tilde G$}}

\def\ga            {\mbox{$g_a$}}
\def\gam           {\mbox{$(g_a)^{-1}$}}
\def\gb            {\mbox{$g_b$}}
\def\gbm           {\mbox{$(g_b)^{-1}$}}

\def\Gm            {G}

\def\gO            {\mbox{$\brev{\g}$}}
\def\gOm           {\brev{\g}}

\def\goheins       {\hat\g_\circ^{(0)}}

\def\gosec         {\tilde{\bar\g}_0}

\def\h             {{\sf h}}

\def\hil           {\mbox{$\calH$}}

\def\hsa           {horizontal subalgebra}

\def\htil          {\tilde h}

\def\hy            {$\mbox{-\hspace{-.66 mm}-}$}

\def\ii            {{\rm i}}

\def\iN            {\!\in\!}

\def\irrep         {irreducible representation}
\def\J             {J}

\def\kma           {Kac\hy Moo\-dy algebra}

\long\def\labl#1   {\label{#1}\ee \ifnum\draftcontrol=1
                   \mbox{ }\\[-12 mm]\query{#1}\\[5 mm] \fi}

\def\lie           {Lie algebra}
\def\Lie           {Lie group}

\def\lV            {\bar L^{\scriptscriptstyle\vee}}
\def\lw            {\mbox{$\overline L^{\rm w \Vee}$}}

\def\modinv        {modular invarian}

\def\nontriv       {non-trivial}
\def\nscon         {non-simply connected}

\def\oaii          {\sum_{l=0}^{N_i-1}\!a_{i,\omD^li}}
\def\oaij          {\sum_{l=0}^{N_j-1}\!a_{i,\omD^lj}}

\def\olie          {orbit Lie algebra}

\newcommand\omb   {\bar\omega}
\newcommand\ombsec   {\tilde{\bar\omega}}

\newcommand\omd[1] {{\dot\omega #1}}
\def\omD           {\dot\omega}

\def\parz          {\frac\partial{\partial z}}
\def\parzb         {\frac\partial{\partial \bar z}}

\def\Pro           {\mbox{\sc p}^{}_{\!\omega}}
\def\Prosec        {\tilde{\mbox{\sc p}}^{}_{\!\omega}}

\def\proj          {\mbox{\sc p}^{\star-1}_{\!\omega}}

\def\Prom          {\mbox{\sc p}^{-1}_{\!\omega}}
\def\Promsec       {\tilde{\mbox{\sc p}}^{-1}_{\!\omega}}

\def\pv            {p^\Vee}

\long\def\query#1{\hskip 0pt{\vadjust{\everypar={}\small\vtop to 0pt{\hbox{}%
     \vskip -13pt\rlap{\hbox to 50.0pc{\hfil{\vtop{\hsize=8pc\tolerance=6000%
     \hfuzz=.5pc\rightskip=0pt plus 3em\noindent#1}}}}\vss}}}}%
\def\range         {{\rm range}}

\def\rep           {representation}
\def\Rep           {Representation}

\def\rmd           {{\rm d}}

\newcommand\sect[1] {\section{#1}\setcounter{equation}{0}}
\newcommand\Sect[2]{\sect{#1}\label{s.#2} \ifnum\draftcontrol=1 \query{s.#2}\fi}

\def\tauo          {{\tau_\omega}}

\def\tsec          {{\tilde T}}

\def\twodim        {two-di\-men\-si\-o\-nal}

\def\Vee           {{\scriptscriptstyle\vee}}

\newcommand\version[1] {\ifnum\draftcontrol=1 \typeout{}\typeout{#1}\typeout{}
                   \vskip3mm \centerline{\fbox{\tt DRAFT -- #1 -- \today}}  
                   \vskip3mm \fi}

\def\Wbhat         {\hat{\bar W}}

\def\wh            {\hat w}

\def\What          {\hat W}

\def\WO            {\brev{W}}

\def\wsp           {w_0}
\def\WZW           {Wess\hy Zumino\hy Witten}
\def\wzwm          {WZW model}
\def\wzwt          {WZW theory}
\def\wzwts         {WZW theories}

\def\zet           {{\dl Z}}

\def\zetplus       {\mbox{${\zet}_{>0}$}}

\def\dl {\bf }

\global\def\draftcontrol{0}
\catcode`\@=12

\documentstyle[12pt]{article}

\begin{document}\version\versionno
%%% for draft versions, suppress in definitive version
%\draft

\begin{flushright}  {~} \\[-23 mm] {\sf hep-th/9611092} \\
{\sf UCB-PTH-96/53} \\ {\sf LBNL-39585} 
\\[1 mm]{\sf November 1996} \end{flushright} \vskip 2mm

%%%%%%%%%%%%%%%%%%%%%%%%%%%%%%%%%%%%%%%%%%%%%%%%%%%%%%%%%%%%%%%%%%%%%%%
\begin{center} \vskip 12mm

{\Large\bf ON MODULI SPACES OF FLAT CONNECTIONS } \vskip 0.3cm
{\Large\bf WITH NON-SIMPLY CONNECTED} \vskip 0.3cm
{\Large\bf STRUCTURE GROUP} \vskip 12mm
{\large Christoph Schweigert}\\[5mm] {\small Department of Physics,
University of California, Berkeley, CA 94720, USA}\\
{\small and} \\
{\small Theoretical Physics Group, Lawrence Berkeley National Laboratory}\\ 
{\small Berkeley, CA 94720, USA}
\end{center}

\vskip 15mm

\begin{quote} {\bf Abstract.} \\
We consider the moduli space of flat $\G$-bundles over the \twodim\ torus, where
$\G$ is a real, compact, simple Lie group which is not simply connected. We 
show that the connected components that describe topologically \nontriv\ 
bundles are isomorphic as symplectic spaces to moduli spaces of topologically 
trivial bundles with a different structure group. Some physical applications 
of this isomorphism which allows to trade topological non-triviality for a 
change of the gauge group are sketched. 
\end{quote}

\vskip 12mm

\sect{Introduction}

In this letter we present an isomorphism between two different moduli spaces of 
gauge equivalent classes of flat connections on principal bundles over the 
\twodim\ torus $\Sigma$. Moduli spaces of flat connections over complex curves
have been the subject of intensive investigations, since
they play a key role both in \CST\ in three dimensions and in \twodim\ \cft.
In \CST\ with structure group $\G$ the moduli space $\calmg$ parametrizes the 
space of inequivalent 
classical solutions. Holomorphic quantization allows to associate a \findim\ 
complex vector space to $\calmg$, the space of conformal blocks 
\cite{axdw,witt27}. Appropriate sesquilinear combinations
of elements of these spaces describe the correlation functions of the \wzwm\ 
based on \G.
The latter constitute an important subclass of \twodim\ \cfts; they also
serve as the building blocks of many other \cfts, e.g.\ via the coset 
construction. 

In the present note, we consider the case when the structure group
$\G$ is a compact, connected, finite-dimensional \Lie, which is not simply 
connected. The corresponding moduli spaces arise naturally, e.g.\ in the 
description of \wzwm s or \CSTs\ based on these groups. Another important 
application of these spaces is 
the resolution of field identification fixed points in \cfts \cite{fusS4}. In 
the algebraic approach, the solution of this problem has given rise to a 
surprisingly
rich structure, both in the case of coset \cfts\ \cite{fusS4} and integer spin
simple current modular invariants \cite{fusS6}. It has been argued \cite{hori} 
that in order to describe the resolution
of field identification fixed points in these models in a Lagrangean framework,
\nscon\ structure groups $\G$ have to be considered. The result of this
letter is therefore a first step towards a geometric understanding of the
results of \cite{fusS4,fusS6}.

If the structure group $\G$ is \nscon, the moduli space $\calmg$ consists of
different connected components, which typically have different dimensions. 
Writing \G\ as the quotient of $\Gu$, the universal covering group
of $\G$, by a subgroup $Z$ of the center of $\Gu$, 
\be \G \cong  \Gu / Z \, , \ee
the connected components of $\calm_G$ are labeled by the finite abelian group 
$Z$:
\be \calmg = \bigcup_{\omega\in Z} \calmomega \, . \ee
If $\omega\in Z$ is not the identity, $\calmomega$ is said to describe a 
topologically \nontriv\ sector of the theory. 

The main result of this note is that the moduli space $\calmomega$ 
describing a topologically \nontriv\ sector is isomorphic 
to the moduli space for some {\em other} \Lie\ $\G^\omega$, which
describes the topologically {\em trivial} sector:
\be \calmomega \cong {\cal M}^{\bfem}_{\Gm^\omega} \, . \labl{claim}
The \Lie\ $\G^\omega$ is again simple, \findim\ and compact. 
This isomorphism allows us to trade 
topological non-triviality for some other structure group and to reduce 
calculations to calculations in the topologically trivial sector only.
The moduli spaces $\calmg$ can be obtained as a symplectic 
quotient of the infinite-dimensional symplectic space of all gauge potentials;
as a consequence, a smooth dense open subset of them is a symplectic 
manifold with a symplectic form $\Omega$. The symplectic structure $\Omega$ 
plays an important role, in particular for holomorphic quantization; we will 
see that the isomorphism \erf{claim} respects $\Omega$.

Before we describe how the \Lie\ $\G^\omega$ is obtained from $\omega$ and
$\G$, it is helpful to discuss the implications of this result for the
quantized \CST. To apply the method of holomorphic quantization to 
the spaces ${\cal M}^{\bfem}_{G}$ of topologically trivial connections one picks 
a complex structure on the torus $\Sigma$, parametrized by some complex number 
$\tau$ in the complex upper half plane. This turns $\Sigma$ into a complex 
surface $\Sigma_\tau$, and also induces a complex structure on $\calmg$.
Next, one chooses a holomorphic line bundle $\cal L$ over 
${\cal M}^{\bfem}_{G}$ such that its curvature is given by $2\pi\ii \Omega$,
where $\Omega$ is the symplectic form on  ${\cal M}^{\bfem}_{G}$. 
After fixing a positive integral value, the level $k$ (which plays the role
of a coupling constant for the field theory),
the quantization $\block$ of ${\cal M}^{\bfem}_{G}$ is obtained as the 
\findim\ vector
space of holomorphic sections of the $k$-th tensor power of $\cal L$: 
\be \block := H^0({\cal M}^{\bfem}_{G}, {\cal L}^{\otimes k}) \, . \labl{qspace}

If the underlying Riemann surface is a torus $\Sigma_\tau$, there is a 
distinguished basis for $\block$: 
denote the \lie\ of $\G$ by $\gbar$, and consider the
untwisted affine \lie\ $\g=\gbar^{(1)}$ based on $\gbar$. For fixed level $k$,
there are finitely many unitarizable irreducible highest weight \rep s
$\hil_\Lambda$ of \g. The character 
\be \chi_\Lambda(\tau, h) := {\rm Tr}_{\calH_\Lambda} \eE^{2\pi\ii\tau(L_0 
- c/24)} \eE^{\ii h} \, , \labl{15}
of any \irrep\ is a function depending on $\tau$ and a variable $h$ that takes 
values in the \csa\ $\gbar_0$ of $\gbar$. (After a choice of basis
in $\gbar_0$, $h$ can be described by its components, the so-called Cartan 
angles.)
The characters of all \irrep s at level $k$ form a basis of the space $\block$
\cite{axdw,emss}.

Let us now turn to the case of our interest, when $\G$ is 
not simply connected. The center of the universal covering group
$\Gu$ can be identified with a subgroup of the symmetries of the \dyd\ of the 
affine \lie\ \g. Any symmetry $\omD$ of order $N$ of the Dynkin diagram of $\g$ 
induces an automorphism $\omega$ of the affine \lie\
which acts like $\omega(E^i_\pm) :=  E^{\omd i}_\pm$ on the step operators 
corresponding to the simple roots and $\omega(H^i) := H^{\omd i}$ for the 
generators of the \csa. (The action on a full basis of the centrally extended
loop algebra can be found in section 6 of \cite{fusS3}.) The automorphism 
$\omega$ preserves the
triangular decomposition of $\g$ and in particular the \csa; hence the dual map
$\omega^\star$ restricts to an isomorphism of the weight space $\g_0^\star$
of $\g$.

The automorphism $\omega$ gives rise \cite{fusS3} to `twisted intertwiner maps',
i.e.\ linear maps between irreducible highest weight \rep s of \g\ 
\be \tauo \, :\quad \hil_\Lambda \to \hil_{\omega^\star(\Lambda)} \ee
which obey 
\be \tauo \, x = \omega(x) \, \tauo \quad\mbox{for all} \, \, x\in \g\,   \ee
and which map the highest weight vector of $\hil_\Lambda$ to the highest weight
vector of $\hil_{\omega^\star(\Lambda)}$. Those weights for which
$\omega^\star(\Lambda)=\Lambda$, so-called {\sl fixed points} of 
$\omega^\star$, are of particular interest: in this case $\tauo$ is an 
endomorphism, and one can 
insert $\tauo$ in the trace \erf{15} to obtain a new set of functions on the
\csa, the so-called {\sl twining characters}
\be \chi^\omega_\Lambda(\tau,h) 
:= {\rm Tr}_{\calH_\Lambda} \tauo \eE^{2\pi\ii\tau(L_0 - c/24)} \eE^{\ii h} 
\, . \ee
The twining characters are dominated by the ordinary characters, 
and hence they converge wherever the ordinary characters converge. 
We will be interested in those symmetries of \g\ which are associated to
elements of the center of $\Gu$; these describe \cite{scya6} 
the action of a so-called simple current in the corresponding \wzwt. 
The twining characters are close relatives of the index in a supersymmetric
theory, where $(-1)^F$, $F$ the fermion number, plays the role of $\tauo$.
Indeed, in any rational superconformal field theory the supercurrent is a simple
current.

It was shown in \cite{fusS3} (see also \cite{furs}) that the twining character
is identical, in a sense to be made precise later, to the character of some 
other \lie, the {\sl \olie} $\gO$. (The \olie\ depends on both \g\ and 
$\omega$; for the ease of notation, we do not mark the dependence on 
$\omega$ explicitly.)
The \dyd\ of this \lie\ is obtained by
folding the \dyd\ of \g\ according to the symmetry $\omD$. More explicitly,
the Cartan matrix of $\gO$ is defined as follows:
Denote the Cartan matrix of $\g$ by $A=(a_{ij})_{i,j\in I}$,
where the index set is $I=\{0,1,\ldots, {\rm rank} \, \gbar\}$. The symmetry
$\omD$ of the \dyd\ organizes $I$ into orbits of different length $N_i$; we 
choose a set $\hat I$ of representatives in $I$ from each $\omD$-orbit. 
The Cartan matrix of the \olie\
is then labelled by the subset $\cI$ of the set of orbits $\hat I$:
  \be \cI:=\{i\iN{\hat I}\mid\oaii > 0 \} \,. \ee
For any orbit we denote by $s_i$ the number
 \be s_i:=\cases{{a_{ii}/\oaii}\,,&if $i\iN\cI$ and $a_{ii}\not=0\,,$\cr
  \noalign{\vskip 2pt} \ 1\,,&otherwise$\,$,\cr}  \ee
which is either $1$ or $2$. The elements of the Cartan matrix
${\brev A}=({\caij})_{{i,j}\in {\brev I}}$ of the orbit \lie\ $\gO$ are then 
given by 
  \be \caij:=s_j\oaij. \ee
Note that $\cI$ can be the empty set, in which case the \olie\ is the trivial 
\lie.  One can show that the \olie\ of an affine \lie\ is again an 
affine \lie, unless it is trivial. 
\futnote{ All results on \olie s and twining characters are valid in a much 
more general context: they hold for arbitrary generalized \kma s \cite{furs}.}

We emphasize that the \olie\ $\gO$ is not constructed as a subalgebra of \g; 
in particular, the \olie\ is in general not isomorphic to the subalgebra of 
\g\ that is fixed under $\omega$. There is however a natural map $\Pro$ from 
the subspace $\g_0^{(0)}$ of the \csa\ that is fixed under $\omega$ to the 
\csa\ $\gO_0$
of the \olie\ \cite{fusS3}. It is a bijection and the  invariant bilinear 
forms on $\g_0^{(0)}$ and $\gO_0$ are related by 
  \be (h\mid h') = \frac 1N\, (\Pro(h)\mid \Pro(h')) \, , \labl{normrel}
for all $h,h'\in \goheins$. (Recall that $N$ is the order of $\omega$.)
The dual relation for weights reads
  \be  (\lambda\mid\mu) = N\cdot (\proj(\lambda)\mid \proj(\mu)) \, , \labl{nrw}
where we have assumed that the invariant bilinear form on weight space
is normalized such that the highest root of the horizontal subalgebra
has length squared 2. With this notation
the statement that the twining characters are given by the characters
of the \olie s can be made precise \cite{furs,fusS3}:
  \be \Chil(\tau,h) = \ChiO_{\proj(\Lambda)}(\tau,\Pro(h)) \, . \labl{thm}
Our results show that this theorem and the isomorphism \erf{claim} are 
closely related by the procedure of holomorphic quantization.

The key property of the \olie\ $\gO$ is as follows: the dual map $\omega^\star$
acts as a linear map on the weight space of $\g$, on which also
the Weyl group $W$ of $\g$ acts. It can
be shown \cite{furs} that the subgroup $\What$ of $W$ that consists of all
elements of $W$ that commute with $\omega^\star$ is isomorphic to the 
Weyl group
$\WO$ of the \olie\ $\gO$. This fact enters crucially in the proof of 
\erf{thm}; it will also be used in the present letter.
Indeed, the simply connected \Lie\ $\G^\omega$ appearing in \erf{claim} is 
just the simply connected compact \Lie\ whose \lie\ is the horizontal 
subalgebra of $\gO$. 

The rest of this letter is organized as follows: in Section 2 we derive an 
explicit
description of the moduli space $\calmomega$, which is used to set up a map
realizing the isomorphism \erf{claim}. In Section 3 we check that this map
preserves the symplectic structure and derive a condition on the level which
is necessary for the existence of a quantization. In the last section we
comment on applications of our result and present the conclusions.

\sect{The isomorphism}

In this section we will derive an explicit description of $\calmomega$. To
this end we use 
the description of moduli spaces of flat connections
in terms of monodromies around \nontriv\ cycles: the group $\G$ acts on
the space of all group homomorphisms from the fundamental
group $\pi_1(\Sigma)$ to $\G$ by conjugation. This action is just the action 
of gauge symmetries on the monodromies; the moduli space is then isomorphic
to the quotient ${\rm Hom} (\pi_1(\Sigma), \G) / \G $.

The fundamental group of the torus is $\zet^2=\zet\times\zet$, and for
simply connected $\G$ we have to classify all solutions of the equation
\be \ga \gb \gam \gbm = \bfe \, .  \labl{11}
for $\ga,\gb\in\G$, up to a simultaneous conjugation of $\ga$ and $\gb$.
In the case of \nscon\ $\G$ we prefer to work with elements of the universal
covering group $\Gu$ rather than with elements of $\G \cong \Gu/Z$. Thus we 
have to find all solutions $(\ga,\gb)$, with $\ga,\gb\in \Gu$, of the equation
\be \ga \gb \gam \gbm = \omega  \, , \labl1
where $\omega\in Z$ labels the topological sector. Again we have to 
identify solutions that are related by a simultaneous conjugation 
with some element of $\Gu$. 

In the topologically trivial sector equation \erf1 tells us that $\ga$ and
$\gb$ commute. For any two commuting elements of the real compact \Lie\ $\Gu$ 
there is a maximal torus
containing both elements. The maximal torus is isomorphic to the \csa\
divided by the coroot lattice $\lV$; the intersection of the orbits 
of conjugation with a maximal torus $T$ are just the orbits of the 
Weyl group $\barW_T$.
As a consequence, the moduli space in the topologically trivial sector is 
\be \calmg^{\bfem} \cong  (\gbar_0/\lV \times \gbar_0/\lV ) / \barW_T \, , 
\labl{standard}
where $\barW_T$ acts diagonally.

The analogous analysis in the case of topologically \nontriv\ sectors is
more involved; for the ease of the reader we present the result immediately. 
Consider the untwisted affine \lie\ $\g=\bar \g^{(1)}$ with horizontal 
subalgebra $\bar\g$. To any element $\omega$ of the center of $\Gu$ is 
associated a diagram automorphism of 
$\g$ corresponding to a simple current. (These correspond to the
symmetries of the
\dyd\ of \g\ which are not already symmetries of the \dyd\ of $\gbar$.) 
Upon identifying the \lie\ $\h_T$ of $T$ (which is a \csa\ for the
\lie\ $\gbar$ of $\Gg$) and
the horizontal projection of the \csa\ of $\g^{(1)}$, $\omega$ gives rise to
an affine map on $\h_T$. This map can be expressed in terms of an 
element $w_0$ of the Weyl group $\barW_T$ of $\gbar$ and a shift by an 
element $\pv$ of the coweight lattice 
\be \omb(h) = \pv + w_0(h) = \pv+  \aw h \aw^{-1} \quad . \labl2
Here we have chosen a group element $\aw$ in $\Gu$ to implement the action
of the Weyl group element $w_0$; the element $\aw$ is only determined up to 
an element of the maximal torus $T$, and we will have to fix some convenient
choice for $\aw$.  The 
map $\omb$ leaves the (horizontal projection of the) fundamental affine Weyl 
chamber invariant. We will show that 
any solution of equation \erf1 is conjugate to a solution of the form
\be \ga = \exp(\ii h) \aw  \qquad\mbox{and}\qquad \gb = \exp(\ii (h_0+ h') )
\, \labl{res}
where $h_0,h$ and $h'$ are elements of the \csa\ $\h_T$ that obey
$w_0(h) = h$, $w_0(h')=h'$ and $\omb(h_0)= h_0$, respectively. 
Using the map $\Pro$ the elements $h$ and $h'$ can be identified
with elements of the \csa\ of the \olie\ \gO. We are interested in group
elements and therefore any two solutions for which $h$ and $h'$ 
differ by elements of the coroot lattice should be identified. We will see 
that the elements $\beta^\Vee$
of the coroot lattice of $\gbar$ respecting the conditions 
$w_0(\beta^\Vee)=\beta^\Vee$ are 
in one-to-one correspondence to elements of the coroot lattice of 
$\overline{\gO}$. Next,
we also have to take into account the effect of simultaneous conjugation with
elements of $\Gu$ that preserve the conditions on $h$ and $h'$. We will see
that this is described by the diagonal action of the subgroup $\Wbhat$
of the Weyl group $\barW$ of $\gbar$ that commutes with $\omb$. This subgroup, 
however, is isomorphic to the Weyl group of the \hsa\ of the
\olie, and comparing with \erf{standard} we obtain the isomorphism \erf{claim}.

\bigskip
In order to prove that any solution of \erf1 is indeed conjugate to 
\erf{res}, we fix a maximal torus $T$ of $\Gg$ 
that contains $\gb$ (for any element of a compact real \Lie\ such a torus 
exists); we can then write
\be \gb = \exp(\ii h'') \,  \ee
with $h''\in \h_T$. The element $\omega$ of the center can be written as 
$\omega=\exp(-\ii\pv)$, where $\pv\in \h_T$ is an element of the co-weight 
lattice $\lw$ of \Gg\ relative to $T$. Moreover, by adding elements of
the co-root lattice and after choosing the convention for dividing the roots 
into positive and negative roots appropriately,
we can assume that $h''$ is an element of the (closure of the) fundamental 
affine Weyl chamber. Without loss of generality 
we can write $\ga$ as
\be        \ga= r \aw     \, , \ee
where $r$ is some element in \Gg. At the present stage, we will fix some 
arbitrary choice for $\aw$; later on, we will determine a canonical choice 
for $\aw$.

Equation \erf1 then becomes
\be \gb \exp(-\ii \pv) = r \aw \exp(\ii h'') \aw^{-1} r^{-1} 
= r \exp(\ii \wsp(h'')) r^{-1} 
= r \exp(\omb (\ii h'')) \exp(-\ii \pv)  r^{-1} \, .  \labl{pp}
The element $\omega = \exp(-\ii \pv)$ is in the center of $\Gu$, 
equation \erf{pp} is therefore equivalent to
\be \exp(\omb(\ii h'')) = r^{-1} \exp(\ii h'') r \, . \labl{commut}
We now observe that $\omb$ preserves the \csa\ of $\gbar$, hence the left hand
side is in the maximal torus again. Since the orbits of conjugation on the 
maximal torus are the orbits of the Weyl group $\barW_T$, we find that the 
right hand side is equal to $\exp(\ii w(h''))$ with $w$ a suitable element of 
the Weyl group $\barW_T$ of $\gbar$. However, $\omb$ preserves the fundamental 
affine Weyl chamber, and the 
only Weyl group element that does the same is the identity. 
{}From this we learn that $\omb$ leaves $h''$ fixed, $\omb(h'')=h''$, and as a
consequence, \erf{commut} shows that $r$ and $\gb=\exp(\ii h'')$ commute. Now 
any two commuting elements of the real compact \Lie\ $\Gg$ are contained in 
some maximal torus $\tilde T$, and since all maximal tori of $\Gg$ are 
conjugate, we can find $g\in\Gg$ such that
\be \tilde T = g^{-1} T g \, . \ee
Denoting by $\h_{\tilde T}$ the \csa\ of $\gbar$ that is the \lie\ of
$\tilde T$, we have
\be \gb = g \exp(\ii\eta'') g^{-1} \quad\mbox{and}\quad r = g \exp(\ii\eta) 
g^{-1} \ee
with $\eta,\eta''\in \h_{\tilde T}$. This allows us to rewrite $\ga$ as
\be \ga = r \aw  = g (\eE^{\ii\eta} g^{-1} \aw g ) g^{-1} \, . \ee
Notice that $\tilde\aw := g^{-1} \aw g$ represents the element of the Weyl
group $\barW_{\tilde T}$ for the new maximal torus $\tilde T$ that corresponds 
to the same abstract Weyl group element as the one in $\barW_T$ described by 
$\aw$.

Since we are interested in solutions of \erf1 only up to conjugation, we can
drop the tildes and find that any solution of \erf1 is conjugate to
\be \ga = \exp(\ii h) \aw \quad\mbox{and}\quad \gb = \exp(\ii h'') ,\labl{110}
where $\exp(\ii h)$ and $\exp(\ii h'')$ are elements of the same maximal 
torus $T$ of $\Gg$, and $h'$ is in the fundamental Weyl chamber. By 
conjugation with $g$, $\omb$ gives also rise to an analogous affine map on 
the \lie\ of the new maximal 
torus; $h''$ is invariant under the analogue of $\omb$ on the new maximal torus.

The space $\aff$ of $\omb$-invariant elements of the \csa\ is an affine space
relative to the vector space $\calf:= \ker(1-\wsp)$. 
The map $\Pro$ gives an isomorphism
between the affine space $\aff$ and the horizontal projection of the
weight space of the \olie; moreover, it also provides us with a distinguished 
base point in $\aff$: $h_0 := \Prom(0)$; we write
\be \gb = \exp(h_0 + h')\, , \quad\mbox{where} 
\quad h'\in\calf=\ker(1-\wsp) \, . \ee

We now have now 
to give a more detailed description of the element $h\in \gbar_0$ 
in \erf{110}: we remark that simultaneous conjugation of $\ga$ and 
$\gb$ with an element $\exp(\ii\htil)$ of the maximal torus does not change
$\gb$. The action on $\ga$ can be computed as follows:
\be \begin{array}{lll}
\ga=\exp(\ii h) \aw & \mapsto & \exp(\ii \htil) \exp(\ii h) \aw 
\exp(-\ii \htil)     \\[1em]
              &=& \exp(\ii (h+\htil - \wsp(\htil))) \aw  \, . 
\end{array}\ee

Hence we always change $h$ by a conjugation to 
$h \mapsto h + \htil - \wsp(\htil)$ and obtain an equivalent solution, i.e.\
we are free to add elements of the subspace
$\range(1-w_0)$. Using the fact that $w_0$ is an orthogonal transformation,
this subspace can be expressed as 
\be 
\range(1-w) = (\ker (1-\wsp^t) )^\perp = (\ker (1-\wsp^{-1}) )^\perp 
            = (\ker (1-\wsp))^\perp  \, . \ee
Hence we can assume, after conjugating $\ga$ and $\gb$ simultaneously with a
suitable element of the form $\exp(\ii\htil)$, that both $h$ and $h'$ are 
in the kernel $\calf$.

We have shown that the solutions 
of \erf2 are all conjugated to a solution of the form 
\be \ga= \exp(\ii h) \aw \quad\mbox{and}\quad \gb = \exp(\ii(h_0 + h')) 
= \exp(\ii h_0) \exp(\ii h') \, \labl{cond}
where $h$ and $h'$ are in $\calf$, and $h_0 + h'$ is in the (horizontal
projection of the) fundamental affine
Weyl chamber. Conversely, it is easy to check that any such pair of elements
gives indeed a solution of \erf1.

At this point it seems as if there were an asymmetry between $\ga$ and $\gb$.
However, the situation is indeed symmetric: both $h$ and $h'$ are in 
$\calf$, and hence $\aw$ commutes with $\exp(\ii h)$ and $\exp(\ii h')$. 
This shows that we
can find a second maximal torus $\tsec$, which contains $\aw$, $\exp(\ii h)$ 
and $\exp(\ii h')$, but not $\exp(\ii h_0)$. (Note that $h$ and $h'$ are fixed  
under $\wsp$ and therefore so-called singular elements of the \csa;
hence their exponentials can be indeed contained in two different maximal 
tori.) 

The trivial rewriting of \erf1
\be \gb \ga \gbm \gam = \omega^{-1}  \, , \ee
allows us to change the roles of $\ga$ and $\gb$ in the above considerations,
provided we replace $\wsp$ by $\wsp^{-1}$ and $\omb$ by $\omb^{-1}$. 
In particular, we can write $\ga= \exp(\ii h) \aw$ as the exponential of some
element $\eta$ of the \csa\ $\gosec$ belonging to $\tsec$, $\ga=\exp(\ii \eta)$,
where $\eta$ is invariant under the map $\ombsec$ acting on $\gosec$.
Again, there is a natural base point for this affine space: 
$\eta_0 := \Promsec(0)$, where $\Promsec$ is the analogue of $\Prom$ for 
$\tilde T$. Now recall that the element $\aw$ implementing the Weyl 
group transformations is only specified up to an element of the maximal 
torus $T$. The group element $\exp(\ii \eta_0)$ differs from $\aw$
only by an element of the form $\exp(\ii \tilde h)$ with $\tilde h\in\calf$, 
which is an element in the intersection of $T$ and $\tsec$. Hence we are 
free to replace $\aw$ by $\exp(\ii \eta_0)$ for our considerations.

Having found two distinguished base points, we can now describe any
solution of the form \erf{cond} in a natural way in terms of the \olie: 
use the isomorphisms $\Pro$ and $\Prosec$ to associate to it the pair 
$(\Pro(h_0+h), \Prosec(\eta_0+h')) $ in the weight space $\affO\times\affO$ 
of the \olie.

We are only interested in group elements rather than \lie\ elements, and
we should identify solution of \erf{cond} for which $h$ and $h'$ differ
by elements of the coroot-lattice. However, we must preserve the condition
that $h$ and $h'$ are fixed under $\wsp$; so we are only allowed to add
elements $\beta^\Vee$ of the coroot lattice that are fixed under $\wsp$ 
themselves. Then the translation by $\beta^\Vee$ is an element of the affine
Weyl group of $\g$ that commutes with $\omb$; hence it is in the subgroup 
$\What$
and corresponds to an element in the affine Weyl group of the \olie\ which is
a translation by a coroot of $\bar{\gO}$. This shows that, after applying 
$\Pro$ and $\Prosec$, we simply have to project modulo the coroot lattice of 
the \hsa\ of the \olie\ $\gO$:
\be \affO\times \affO \to 
\affO/ {\brev L}^\Vee  \times\affO/ {\brev L}^\Vee \, . \ee

The only freedom we are left with now is simultaneous conjugation of $\ga$ and 
$\gb$ with a group element $g\in\Gu$, such that the relations in \erf{cond} are
preserved: then along with 
$\gb=\exp(\ii (h_0+ h')$ we have $g \gb g^{-1} = \exp(\ii \tilde h') $.
The two elements $h_0+ h'$ and $\tilde h'$ are related by some element of the
Weyl group $\barW$; moreover, both $\tilde h'$ and $h_0+h'$ 
are fixed under 
$\omb$.  It was shown in the proof of proposition
3.3.\ in \cite{furs} that then the Weyl group element $\wh$ relating $h_0+ h'$ 
and $\tilde h'$ can be chosen in $\Wbhat$: 
\be g\exp(\ii (h_0+ h')) g^{-1} = \exp(\ii \wh(h_0+ h')) \, . \labl{hh}
The isomorphism $\Pro$ intertwines the action of $\Wbhat$ and the one of the
affine Weyl group of the \olie. Hence $\Wbhat$ leaves the base point fixed and
therefore 
\be g\exp(\ii (h_0+ h')) g^{-1} = \exp(\ii (h_0+ \wh h') \, . \ee
Analogous considerations can be applied to $\ga$, using this time the other
maximal torus $\tsec$; we reach the analogous conclusions. Combining the
two results, we find that the remaining
redundancies are taken into account by the diagonal action of the Weyl group
$\Wbhat$:
\be \calmomega \cong  ( \affO/ {\brev L}^\Vee  \times\affO/ {\brev L}^\Vee )
/ \Wbhat  \, . \ee

Comparing this explicit description of the moduli space $\calmomega$
with the standard description \erf{standard} of the moduli space of the
orbit theory in the topologically trivial sector, we obtain the isomorphism
\erf{claim}.

\sect{The symplectic structure}

In this section we want to extend the isomorphism \erf{claim} to include 
also the symplectic structure on a smooth open subset of $\calmomega$.
To this end we construct explicitly a topologically \nontriv\ $\G$-bundle 
and a connection on it with the appropriate monodromies. We fix from now
on a complex structure on the torus, which is parametrized by a complex
number $\tau=\tau_1+\ii\tau_2$ with positive imaginary part $\tau_2> 0$. 

The corresponding torus $\Sigma_\tau$ can be obtained as the quotient of 
the complex plane by the following action of $\zet^2=\zet\times\zet$:
\be R(m,n)(z) :=  z+ m + n \tau\, , \, \mbox{where}\quad m,n\in\zet \, . \ee
The principal $\G$-bundle is obtained by extending this action to an action
on the trivial principal $\G$-bundle $\complex\times \G$ over $\G$: we fix an 
element $h_0$ with $\omb(h_0) = h_0$ and set:
\be R_{\Gm}(m,n)(z,g) 
:= (z+ m + n \tau, \, \eE^{\ii h_0 n} \aw^m g) \, . \labl{twist}
Here $g$ is an element of the \nscon\ group $\G$, and we have written for 
simplicity $\aw$ for the projection to $\G$ of the element $\aw\in\Gu$ of the 
universal covering group we considered in the previous section. 
This defines indeed an action of $\zet^2$, as can be seen as follows:
the equation $w_0(h_0) = h_0 - \pv$ implies that one has in $\Gu$
\be \eE^{\ii h_0 n} \aw^m \eE^{\ii h_0 n'} \aw^{m'}
= \eE^{-\ii n' m  \pv} \eE^{\ii h_0 (n+n')} \aw^{m+m'} \, . \ee
The projection of the first element to the \nscon\ group
$\G$ is trivial, and hence we have indeed an action of $\zet^2$. 
$(\complex\times \G)\, / \, \zet^2$ is a topologically non-trivial 
principal $\G$-bundle over $\Sigma_\tau$.

Let us now fix $h,h'\in\gbar_0$ such that $w_0(h)=h$ and $\wsp(h')=h'$. 
Introduce the element
\be u := h' + \tau h \,  \ee
of the complexification of $\gbar_0$; the connection 
\be A(z) :=  \frac1{2\tau_2} ( -\bar u \rmd z + u \rmd \bar z) 
\labl{pot}
on $\complex\times\G$ is then invariant under the induced action of 
$R_{G}(m,n)$: To see this, we remark that $\aw^{-1}\, u \aw = u$
and $\exp(-\ii h_0) \, u \exp(\ii h_0) = u$. 
Hence the induced action of $R(1,n)$ on the connection \erf{pot} is
\be R(1,n) A(z) 
= \aw^{-1} \eE^{- \ii h_0 n}  A(z-1-b\tau) \eE^{\ii h_0 n} \aw 
= A(z) \, .  \ee

The connection \erf{pot} therefore gives rise to a connection on the principal
$\G$-bundle we constructed; this connection is flat.
Let us now verify that this connection reproduces the monodromies $\ga$ and 
$\gb$. We parametrize the first homology cycle $C_a$ 
by $z(t) = t$ with $0\leq t \leq 1$ and see that
\be
\int_{C_a} A = \int_0^1 \rmd t A(\parz + \parzb)  = \frac1{2\tau_2} (u-\bar u)
= \ii h \, . \ee
Taking into account the additional twist by $\aw$ in \erf{twist}, the monodromy
around $C_a$ is indeed $g_a = \exp(\ii h) \aw$. The second generator $C_b$ 
of one-cycles can be parametrized as $z(t) = \tau t$; we find that 
\be \int_{C_b} A = \int_0^1 A(\tau\parz +\bar\tau \parzb)  \
= \frac1{2\tau_2} (\bar\tau u-\tau \bar u) = \ii h' \, . \ee
Taking again into account the additional twist, we see that the monodromy 
around $C_b$ has the correct value $\gb = \exp(\ii (h'+h_0))$ as well.

The symplectic structure is defined on the tangent space, which consists
of $\gbar$-valued one-forms, which we can assume to be of the following form:
define $\delta u := \delta h' + \tau \delta h$ as a complex linear combination
of two elements $\delta h$ and $\delta h'$ of the \csa\ of the compact real 
form. An arbitrary element of the tangent space is then of the form
\be \delta A 
= \frac1{2\tau_2} ( - \delta \bar u \rmd z + \delta u \rmd \bar z)\, , \ee
and the symplectic form is given by 
\be \Omega_\g(\delta A_1, \delta A_2) 
:= \frac12 \int \rmd^2z \, \kappa_{\g}(\delta A_1 \wedge \delta A_2) \, \ee
where $\kappa_{\gbar}(\cdot,\cdot)$ is the Killing form on the \lie\ $\gbar$. 
Again, we adhere to the convention that Killing forms are normalized such that 
the highest $\gbar$-root has length squared two. A standard calculation gives
\be \Omega_\g(\delta A_1, \delta A_2) = - \frac{\ii}2 
(\kappa_{\gbar}(\delta h'_1, \delta h_2) 
- \kappa_{\gbar}(\delta h_1,\delta h'_2) \,)\, , \ee
which shows that the symplectic form is real and
independent of the complex structure, which is parametrized by $\tau$. 

The comparison of this symplectic form with the one on $\calm_{G^\omega}$ 
therefore reduces to a comparison of the Killing forms on \g\ and $\gO$. 
The relation \erf{normrel} shows that they just differ by a factor of $N$, 
where $N$ is the order of $\omega$ in $Z$. Taking into account the level 
$k\in\zetplus$, we see that
\be k \Omega_\g(\delta A_1, \delta A_2) = 
\frac k N \Omega_{\gOm}(\delta \AO_1, \delta \AO_2) \,. \ee
This shows that upon expressing the symplectic form in terms of quantities in
the \olie\ the level is divided by the order of the automorphism;
this is exactly the relation between the levels of the \lie\ and its \olie\ 
that was derived in \cite{fusS3}. An important consequence is that 
$k \Omega_\Gm$
is an element of the {\em integral} cohomology only if the level is a multiple
of $N$. Only in this case the moduli space can be quantized: this is the 
geometric counterpart of the fact that fixed points only occur at levels which
are multiples of the order of the automorphism.

\section{Applications and Conclusion}

In this letter we have proven an isomorphism which, in physical terms, allows
to trade topological non-triviality for a different gauge group. This result
has several applications: the moduli spaces $\calmomega$ appear
naturally in the description of \CSTs\ or WZW-models on \nscon\ group 
manifolds. For the latter theories (indeed, for any integer spin simple 
current extension \cite{scya6} of
a \cft) a formula for the modular matrix $S$ was derived in \cite{fusS6}.
The isomorphism \erf{claim} will be one ingredient to a rigorous proof of
this formula.
This formula in turn give a Verlinde formula for the dimension
of the space of conformal blocks with a \nscon\ structure group, a problem
that recently also has received attention in algebraic geometry \cite{beau2}.

Another application of the isomorphism \erf{claim} are coset
\cfts, in the description as gauged \wzwts: it has been argued \cite{hori} 
that in these theories one actually has to gauge a \nscon\ group. This 
observation has lead to the conjecture that the contributions 
from the topologically \nontriv\ sectors account for the resolution of
field identification fixed points. This resolution can be written \cite{fusS4}
in terms of quantities of the \olie s. The isomorphism \erf{claim} therefore
lends evidence to the conjecture relating orbit theories and topologically
\nontriv\ sectors; it is also a first step towards a 
better understanding of coset \cfts\ in the Lagrangean framework.

We finally mention that our results only concern principal bundles over a 
\twodim\ torus; it would be interesting to unravel the implications of the 
structures we found for Riemann surfaces of higher genus.

%\newpage 
\vskip4em

  \newcommand{\wb}{\,\linebreak[0]} \def\wB {$\,$\wb}
  \newcommand{\Bi}[1]    {\bibitem{#1}}
  \newcommand{\Erra}[3]  {\,[{\em ibid.}\ {#1} ({#2}) {#3}, {\em Erratum}]}
  \newcommand{\BOOK}[4]  {{\em #1\/} ({#2}, {#3} {#4})}
  \newcommand{\vypf}[5]  {{#1} [FS{#2}] ({#3}) {#4}}
  \renewcommand{\J}[5]   {{#1} {#2} ({#3}) {#4} }
  \newcommand{\Prep}[2]  {{\sl #2}, preprint {#1}}
 \def\anop  {Ann.\wb Phys.}
 \def\foph  {Fortschr.\wb Phys.}
 \def\hepa  {Helv.\wb Phys.\wB Acta}
 \def\ijmp  {Int.\wb J.\wb Mod.\wb Phys.\ A}
 \def\jodg  {J.\wb Diff.\wb Geom.}
 \def\jopa  {J.\wb Phys.\ A}
 \def\npbF  {Nucl.\wb Phys.\ B\vypf}
 \def\npbp  {Nucl.\wb Phys.\ B (Proc.\wb Suppl.)}
 \def\nuci  {Nuovo\wB Cim.}
 \def\nupb  {Nucl.\wb Phys.\ B}
 \def\phlb  {Phys.\wb Lett.\ B}
 \def\comp  {Com\-mun.\wb Math.\wb Phys.}
 \def\lemp  {Lett.\wb Math.\wb Phys.}
 \def\phrd  {Phys.\wb Rev.\ D}
 \def\mpla  {Mod.\wb Phys.\wb Lett.\ A}
 \def\duke  {Duke\wB Math.\wb J.}

 \def\A       {Algebra}
 \def\alg     {algebra}
 \def\Be     {{Berlin}}
 \def\BIR    {{Birk\-h\"au\-ser}}
 \def\Ca     {{Cambridge}}
 \def\CUP    {{Cambridge University Press}}
 \def\furu    {fusion rule}
 \def\GB     {{Gordon and Breach}}
 \newcommand{\inBO}[7]  {in:\ {\em #1}, {#2}\ ({#3}, {#4} {#5}),  p.\ {#6}}
 \def\Infdim  {Infinite-dimensional}
 \def\NY     {{New York}}
 \def\nn      {$N=2$ }
 \def\Q       {Quantum\ }
 \def\qg      {quantum group}
 \def\Rep     {Representation}
 \def\SV     {{Sprin\-ger Verlag}}
 \def\syms    {sym\-me\-tries}
 \def\wzw     {WZW\ }

\bigskip\bigskip
\small
\noindent{\bf Acknowledgement.} \ I am grateful to K.\ Gawedzki for helpful
discussions and to J.\ Fuchs for a critical reading of the manuscript. \\
This work was supported in part by the Director, Office of Energy Research, 
Office of Basic Energy Sciences, of the U.S.\ Department of Energy under
Contract DE-AC03-76F00098 and in part by the National Science Foundation
under grant PHY95-14797.

\end{document}